\documentclass [letterpaper]{IEEEtran}
\usepackage{listings}
\usepackage{xcolor}
\usepackage{tabularx}
\usepackage{comment}
\definecolor{codegreen}{rgb}{0,0.6,0}
\definecolor{codegray}{rgb}{0.5,0.5,0.5}
\definecolor{codepurple}{rgb}{0.58,0,0.82}
\definecolor{backcolour}{rgb}{0.95,0.95,0.92}
\lstdefinestyle{mystyle}{
  backgroundcolor=\color{backcolour}, commentstyle=\color{codegreen},
  keywordstyle=\color{magenta},
  numberstyle=\tiny\color{codegray},
  stringstyle=\color{codepurple},
  basicstyle=\ttfamily\footnotesize,
  breakatwhitespace=false,         
  breaklines=true,                 
  captionpos=b,                    
  keepspaces=true,                 
  numbers=left,                    
  numbersep=5pt,                  
  showspaces=false,                
  showstringspaces=false,
  showtabs=false,                  
  tabsize=2
}
\usepackage{graphicx,
	psfrag,
	epsfig,
	epstopdf,
	amsthm,
	amssymb,
	url,
	subcaption,
	balance,
	enumerate,
	color,
	setspace,
	tikz,
	pgfplots
}
\usepackage{amsmath}
\usepackage[nospace,noadjust]{cite}
\usepackage{pbox}
\usepackage{multirow}
\usepackage{adjustbox}
\usepackage{array}
\usepackage{booktabs}
\usepackage{tabularx}
\usepackage{float}
\usepackage{breqn}
\usepackage{algorithm}
\usepackage{algpseudocode}
\usetikzlibrary{shapes.geometric}

\usepackage{booktabs,siunitx,adjustbox,makecell}
\newcommand{\cref}[1]{Constraint~\ref{#1}}

\newcommand{\ignore}[1]{}
\usepackage[paperwidth=8.5in, paperheight=11in, top=1.05in, bottom=0.95in, left=0.6in, right=0.65in]{geometry}

\begin{document}

%\title{Traffic Management in Congested Areas through a Deep Reinforcement Learning Approach}

%\title{Federation Aware MEC Selection in 5G/B5G Networks}
%\title{Closing the AI Responsibility Gap in Network Management: A Preamble Focus on End-User Rights}
\title{Hybrid Responsible AI-Stochastic Approach for SLA Compliance in Multivendor 6G Networks}
	\author{
	\IEEEauthorblockN{ Emanuel Figetakis, and Ahmed Refaey Hussein}\\

	\IEEEauthorblockA{ University of Guelph, Guelph, Ontario, Canada.}}
	%\IEEEauthorblockA{\IEEEauthorrefmark{2} Western University, London, Ontario, Canada.}}

\maketitle
\begin{abstract}

The convergence of AI and 6G network automation introduces new challenges in maintaining transparency, fairness, and accountability across multivendor management systems. 
Although closed-loop AI orchestration improves adaptability and self-optimization, it also creates a responsibility gap, where violations of SLAs cannot be causally attributed to specific agents or vendors. 
This paper presents a hybrid responsible AI-stochastic learning framework that embeds fairness, robustness, and auditability directly into the network control loop. 
The framework integrates RAI games with stochastic optimization, enabling dynamic adversarial reweighting and probabilistic exploration across heterogeneous vendor domains.  An RAAP continuously records AI-driven decision trajectories and produces dual accountability reports: user-level SLA summaries and operator-level responsibility analytics. 
Experimental evaluations on synthetic two-class multigroup datasets demonstrate that the proposed hybrid model improves the accuracy of the worst group by up to 10.5\%.
Specifically, hybrid RAI achieved a WGAcc of 60.5\% and an AvgAcc of 72.7\%, outperforming traditional RAI-GA (50.0\%) and ERM (21.5\%). The audit mechanism successfully traced 99\% simulated SLA violations to the AI entities responsible, producing both vendor and agent-level accountability indices. These results confirm that the proposed hybrid approach enhances fairness and robustness as well as establishes a concrete accountability framework for autonomous SLA assurance in multivendor 6G networks.

\end{abstract}

\section{Introduction}

The convergence of Artificial Intelligence (AI) and next-generation communication systems has transformed the paradigms of network management \cite{ZeroTouchAI}. 
In emerging 6G architectures, intelligent agents embedded within the Management and Network Orchestration (MANO) layer enable closed-loop automation (CLA) and zero-touch networks (ZTN) for self-healing, self-optimization, and dynamic resource orchestration across domains such as Radio Access Network (RAN), Virtualized Evolved Packet Core (vEPC) and Multi-Access Edge Computing (MEC) \cite{E3, R1, R2}. 
While these mechanisms improve agility and operational efficiency, they also introduce a governance challenge to enforce accountability and Service-Level Agreement (SLA) compliance in multivendor environments where distributed AI systems act autonomously without centralized oversight.

Recent industry frameworks highlight this duality. 
Ericsson’s intent-driven architecture translates high-level policies into self-evaluating control loops \cite{ericsson_2025_intentdriven}, while Nokia’s Sense–Think–Act paradigm envisions proactive management of SLAs through AI-based detection and mitigation \cite{nok_2025_how}. 
Cisco extends these principles to multivendor orchestration with transparent accountability \cite{cisco_2025_closedloop}. 
Although these frameworks automate compliance, they lack unified mechanisms to attribute responsibility or audit AI decisions when SLA violations occur.

This responsibility gap arises from the distributed, opaque nature of multiagent AI systems, each trained on heterogeneous datasets and tuned for local objectives that collectively affect global network behavior. 
Minor misalignments among agents can propagate across control loops, degrading Quality of Experience (QoE) and triggering SLA breaches. 
Conventional tools can detect anomalies, but cannot trace causality or identify which AI entity or vendor action caused a violation. 
Moreover, recent global AI governance regulations \cite{EUAIACT,USAAI,AIGOV-CAD} demand transparency, fairness, and traceability, requirements that cannot be met without runtime interpretability and auditability embedded directly within the AI control plane.

Traditional explainable AI (XAI) techniques such as LIME \cite{ribeiro2016whyitrustyou} and SHAP \cite{lundberg2017unifiedapproachinterpretingmodel} provide post hoc interpretability by correlating input features with model outputs but fail to capture causal responsibility in dynamic, safety-critical environments. 
Indeed, they can explain how a decision was made, but not why it failed or who contributed to the failure. 
Consequently, biases or mispredictions in one subsystem can cascade across vendors, leading to large-scale SLA breaches without traceable accountability. 
Emerging efforts in Causal AI \cite{5Gcausal} and Responsible AI (RAI) \cite{gupta2023responsibleairaigames} advance fairness and transparency but remain static single-agent formulations incapable of modeling the temporal and distributed complexity of 6G orchestration. Building on this gap, our previous work \cite{E1} introduced an Intelligent Audit System that combined Deep Reinforcement Learning and Machine Learning to identify and quantify responsibility among AI-based network management agents, demonstrating the feasibility of measurable accountability in multi-vendor environments. Therefore, accountability in such systems must be both explainable and enforceable, requiring a dynamic RAI-stochastic framework that links every AI decision to its causal origin, contextual intent, and real-time impact on SLAs.

%placed here for the positioning
\begin{figure*}[h]
    \centering
    \includegraphics[width=.98\textwidth,height=82mm]{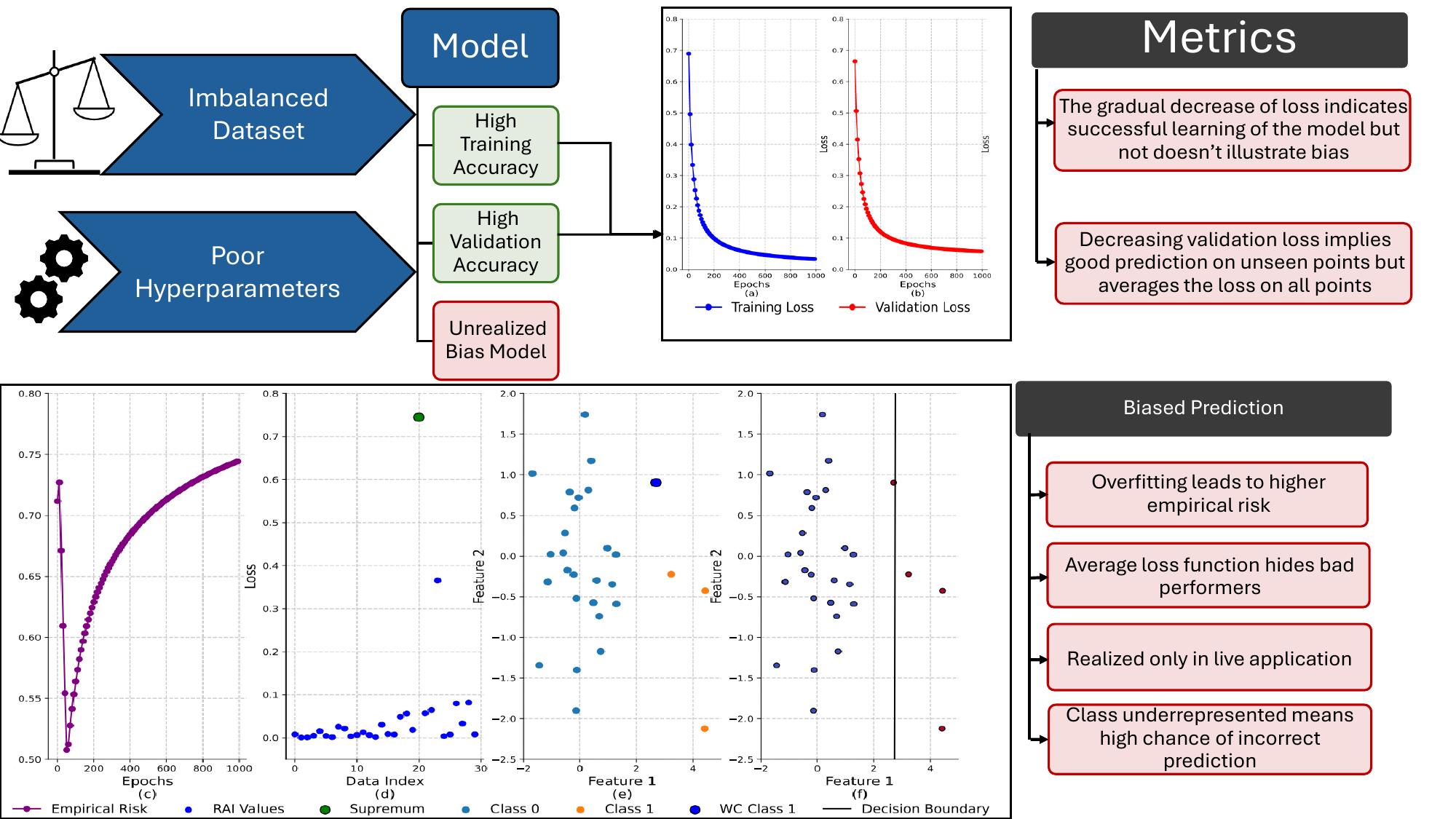}
   % \caption{Imbalanced Logistic Regression Model using Stochastic Gradient Descent.}
    \caption{\textbf{Responsibility-Aware Audit Plane (RAAP):} 
    The RAAP continuously monitors AI-driven control loops within the multi-vendor 6G management stack. 
    It records decision traces $\{(x_t,a_t,s_t,z_t)\}$ and generates two complementary audit views:
    (i) a \textit{user-level SLA dashboard} reporting compliance status and remediation actions, and 
    (ii) an \textit{operator-level accountability report} summarizing average accuracy (AvgAcc), worst-group accuracy (WGAcc), 
    and responsibility attribution per vendor/agent. 
    This design operationalizes the hybrid RAI–stochastic model, ensuring explainable, fair, and auditable SLA compliance.
    }
    \label{fig:LRSGD}
\end{figure*}

To address these challenges, this paper introduces a hybrid responsible AI-stochastic framework that bridges automation and accountability for SLA compliance in autonomous networks. 
The framework integrates RAI games with stochastic optimization to ensure fairness, robustness, and traceability across distributed AI decision pipelines. 
Its main contributions are as follows:
\begin{itemize}
    \item A responsibility-aware audit mechanism that correlates AI decision trajectories with network performance metrics to identify the root causes of SLA violations.
    \item A hybrid RAI–stochastic optimization engine that combines fairness-sensitive learning with probabilistic decision making for adaptive operation under dynamic conditions.
    \item Dual-layer accountability reporting that generates both user-level SLA summaries and operator-level causal responsibility reports.
    \item Quantitative fairness and compliance metrics that extend beyond QoS indicators to evaluate bias, empirical risk, and vendor-agnostic accountability.
\end{itemize}

The remainder of this paper is organized as follows. Section~II presents the Network and Responsibility Modeling Framework. 
Section~III details the Hybrid RAI–Stochastic Learning Framework and its algorithms. 
Section~IV evaluates the proposed method through comparative simulations and compliance metrics, and Section~V concludes with insights into auditable, policy-aligned AI systems for 6G networks.

\section{System Architecture and Accountability Model}

This section formalizes the multivendor 6G management environment underlying the proposed responsibility-aware hybrid learning framework. The system model captures a distributed control architecture in which multiple AI-driven agents operate under service level agreements (SLAs) and heterogeneous vendor domains. The goal is to integrate accountability and fairness directly into the closed-loop decision-making process, thereby enabling explainable, auditable, and reliable network operation through a Responsibility-Aware Audit Plane (RAAP).

The considered 6G management stack consists of two main domains: the Service Domain (SD) and the Management and Orchestration (MANO) domain. The Service Domain includes the physical and virtual network functions managed by $V$ independent vendors, encompassing the radio access network (RAN), the virtualized core, and multi-access edge computing (MEC) components. Each vendor $v \in \{1, \dots, V\}$ controls a set of operational actions $a_t^v \in \mathcal{A}^v$ that affect the global network state $s_t \in \mathcal{S}$, which represents observable quantities such as traffic load, latency, packet loss, and reliability. The network dynamics follow a controlled stochastic process given by
\begin{equation}
\begin{aligned}
s_{t+1} &\sim P(\cdot \mid s_t, a_t),\qquad a_t=(a_t^1,\dots,a_t^V),\\
a_t^v &\in \mathcal{A}^v,\quad s_t\in\mathcal{S}.
\end{aligned}
\end{equation}
where $P$ denotes the transition kernel that governs the evolution of the system under the joint actions of all vendors.

The MANO domain operates as an intelligent control layer that coordinates these distributed entities. It hosts a set of learning agents $\{h_\theta\}$, each following a Sense–Decide–Act cycle. At every decision epoch $t$, an agent observes the telemetry of the system $x_t$ and determines an action $a_t = h_\theta(x_t)$ according to its learned policy. MANO enforces these actions through intent translation mechanisms and provides policy oversight to ensure cross-domain consistency and SLA compliance. Embedded within the MANO is the Responsibility-Aware Audit Plane (RAAP), which continuously records the decision traces $(x_t, a_t, s_t)$, computes SLA indicators, and assigns operational responsibility to specific agents and vendors.

Service-level objectives (SLOs) are represented as a vector of performance constraints $g(s_t) \in \mathbb{R}^m$, defining targets such as latency, throughput, and reliability. A binary SLA violation indicator $z_t$ is introduced to denote whether the system violates any of these objectives at time $t$, such that
\begin{equation}
    z_t =
    \begin{cases}
        1, & \text{if any SLO constraint is violated},\\
        0, & \text{otherwise}.
    \end{cases}
\end{equation}
The MANO layer maintains a log dataset $\mathcal{L}$ comprising tuples $(x_t, a_t, s_t, z_t, \ell_t, \text{agent\_id}, \text{vendor\_id})$, where $\ell_t = \ell(h_\theta(x_t), y_t)$ represents the instantaneous learning loss between the predicted and desired outcomes. This data set supports both real-time monitoring and offline accountability analysis.

Traffic and service data are partitioned into subgroups $\mathcal{G} = \{G_1, \dots, G_K\}$ to reflect various user categories, locations, or network slices. Each subgroup may experience unique conditions due to spatial heterogeneity or vendor-specific implementations. To capture these disparities, an adversarial reweighting factor $w \in \mathcal{W}$ is introduced to model distributional shifts, bias, or vendor drift, thus serving as a fairness-aware abstraction of environmental variability. 

The RAAP component enables two complementary audit views derived from $\mathcal{L}$. The first is the user-level SLA dashboard, which reports compliance status, detected violations, and corrective actions in a transparent and interpretable way. The second is the operator-level accountability report, which aggregates the accuracy and fairness statistics between agents and vendors, including the AvgAcc, the WGAcc, and the vendor responsibility indices. These reports are used not only for interpretability but also for retraining and policy adjustment within the hybrid RAI–stochastic optimization loop.

In general, the proposed system model establishes a closed accountability loop that connects stochastic control, adversarial robustness, and explainable auditing. It forms the foundation for the methodological developments presented in the following sections, where fairness-aware optimization and hybrid stochastic learning are used to enhance SLA compliance and operational transparency in multivendor 6G networks.

\section{Hybrid RAI–Stochastic Learning Framework}

The proposed hybrid RAI–Stochastic Learning Framework integrates fairness-aware optimization, stochastic decision processes, and adaptive exploration under a unified analytical model. Its goal is to ensure that autonomous agents in multivendor 6G systems preserve both SLA compliance and equitable performance across vendors and subgroups. The framework extends the Responsible AI (RAI) game formulation, based on constrained adversarial reweighting, into a hybrid stochastic design that captures temporal variability, uncertainty, and non-stationary network dynamics.

\subsection{RAI Games}

The RAI games \cite{gupta2023responsibleairaigames} formalize fairness as a minimax optimization problem, where the learner seeks to minimize the empirical risk in the worst case under fairness constraints. The weighted empirical risk of a hypothesis $h$ over a dataset $\{(x_i,y_i)\}_{i=1}^n$ is defined as
\begin{equation}
\hat R^{(n)}_{w}(h) = \frac{1}{n}\sum_{i=1}^{n} w_i\,\ell(h(x_i),y_i)
= \mathbb{E}_{(x,y)\sim \hat P_{w}}[\ell(h(x),y)].
\end{equation}
Here, $w_i$ are importance weights induced by fairness constraints. 
The adversary selects the worst-case reweighting $w\in W_n$ that maximizes the empirical loss, yielding the robust objective:
\begin{equation}
\min_{h\in\mathcal{H}} \ \sup_{w\in W_n} \ \hat R^{(n)}_{w}(h),
\label{eq:robust_training}
\end{equation}
where
\begin{equation}
\hat R^{(n)}_{\mathrm{RAI}}(h) = \sup_{w\in W_n} \hat R^{(n)}_{w}(h).
\label{eq:rai_risk}
\end{equation}
This allows the learner to randomize over hypotheses, transforming the problem into a bilinear zero-sum game:
\begin{equation}
\min_{Q\in\Delta(\mathcal{H})} \ \sup_{w\in W_n} \
\mathbb{E}_{h\sim Q,\ (x,y)\sim \hat P_w}[\ell(h(x),y)].
\label{eq:bilinear_game}
\end{equation}

Different sets of constraints $W_n$ (for example, Worst-Group, CVaR\footnote{\textit{Conditional Value at Risk (CVaR), quantifies the expected loss in the worst $\alpha$\% of cases. It penalizes extreme outcomes, emphasizing fairness and robustness under uncertainty.}}, or $\chi^2$) balance robustness and fairness, and because the objective is linear in both $Q$ and $w$, the game converges to a Nash equilibrium. However, the static adversary structure limits adaptability under dynamic network conditions, an issue addressed by the hybrid stochastic formulation.

\subsection{Hybrid RAI–Stochastic Formulation}

To address temporal variability and uncertainty, the hybrid variant introduces a dynamic adversary and stochastic exploration mechanism as in Algorithm~\ref{alg:RAI_Stochastic}.  
Instead of solving a one-shot minimax problem, it iteratively updates both the learner’s policy and the adversary’s response \cite{E2}.  
At each iteration, the learner samples a hypothesis $h_t$ with probability $\epsilon_t$, allowing exploration beyond deterministic updates and avoiding premature convergence.  
The resulting objective function is expressed as:
\begin{IEEEeqnarray}{rCl}
\min_{\{Q_t\}}\;\max_{\{\pi_t\}}\;
&\frac{1}{T}\sum_{t=1}^{T}&
\mathbb{E}_{w\sim\pi_t}\!\Big[
(1-\epsilon_t)\min_{h\in\mathcal{H}}\hat R^{(n)}_{w}(h)\nonumber\\
&&\hspace{5.2em}
+\;\epsilon_t\,\mathbb{E}_{h\sim Q_t}\hat R^{(n)}_{w}(h)
\Big].
\end{IEEEeqnarray}

This dynamic formulation updates the distribution of the learner's policy $Q_t$ through the Follow-The-Regularized-Leader (FTRL) rule and refines the adversary's strategy $\pi_t$ in each round \cite{hazan2023introductiononlineconvexoptimization}, to ensure convergence under evolving multivendor conditions.

\begin{algorithm}[t]
\caption{Hybrid RAI–Stochastic Game Algorithm}
\label{alg:RAI_Stochastic}
\begin{algorithmic}[1]
\Require Training data $\{(x_i,y_i)\}_{i=1}^{n}$, loss $\ell$, hypothesis set $\mathcal{H}$, exploration rate $\epsilon_t$, learning rate $\eta_t$, step size $\alpha_t$

\State Initialize policy distribution $Q_0$ over $\mathcal H$
\State Initialize adversarial strategy $\pi_0$
\For{$t=1$ to $T$}
    \State \textbf{Stochastic Exploration:} sample $h_t\!\sim\!Q_t$ with prob.\ $\epsilon_t$
    \State Compute worst-case weights:
        \[
        w_t=\arg\max_{w\in W_n}\mathbb{E}_w[\ell(h_t(x),y)]
        \]
    \State \textbf{RAI Update:} update ensemble
        \[
        Q_t=(1-\alpha_t)Q_{t-1}+\alpha_t\delta_{h_t}
        \]
    \State \textbf{Adversary Response:}
        \[
        \pi_t=\arg\max_{\pi}\mathbb{E}_{h\sim Q_t,w\sim\pi}[\ell(h(x),y)]
        \]
    \State \textbf{Policy Improvement:}
        \[
        Q_t\gets Q_t+\eta_t\nabla_Q
        \mathbb{E}_{h\sim Q_t,w\sim\pi_t}[\ell(h(x),y)]
        \]
\EndFor
\State \textbf{Return:} final mixed-strategy ensemble $Q_T$
\end{algorithmic}
\end{algorithm}

The RAI component ensures fairness and accountability by constraining worst-group risk, while the stochastic extension enhances adaptability through probabilistic exploration.  
Their integration enables the model to dynamically balance fairness and robustness, making the hybrid framework particularly effective for SLA-aware orchestration in multi-vendor 6G networks operating under uncertainty.

\section{Experimentation}
A baseline was established to allow for a fair comparison among all versions of the traditional and RAI-based training frameworks. Following the methodology and data generation process in \cite{gupta2023responsibleairaigames}, a two-feature, two-class synthetic dataset with 1000 samples was generated to ensure statistical reliability while maintaining computational efficiency for iterative stochastic experiments. The size of the dataset was selected to balance sample diversity and model interpretability, i.e. large enough to produce representative class overlaps for fairness analysis, yet compact enough to visualize decision boundaries and subgroup effects across models. Overlapping points were intentionally introduced to increase the classification difficulty and reveal how each model handles uncertainty and bias. Subgroups were then extracted from the dataset to assess whether underrepresented regions received equitable treatment or if performance was dominated by majority groups, thus illustrating imbalance and fairness differences between baselines.

\begin{figure}[h]
    \centering
    \includegraphics[width=.48\textwidth]{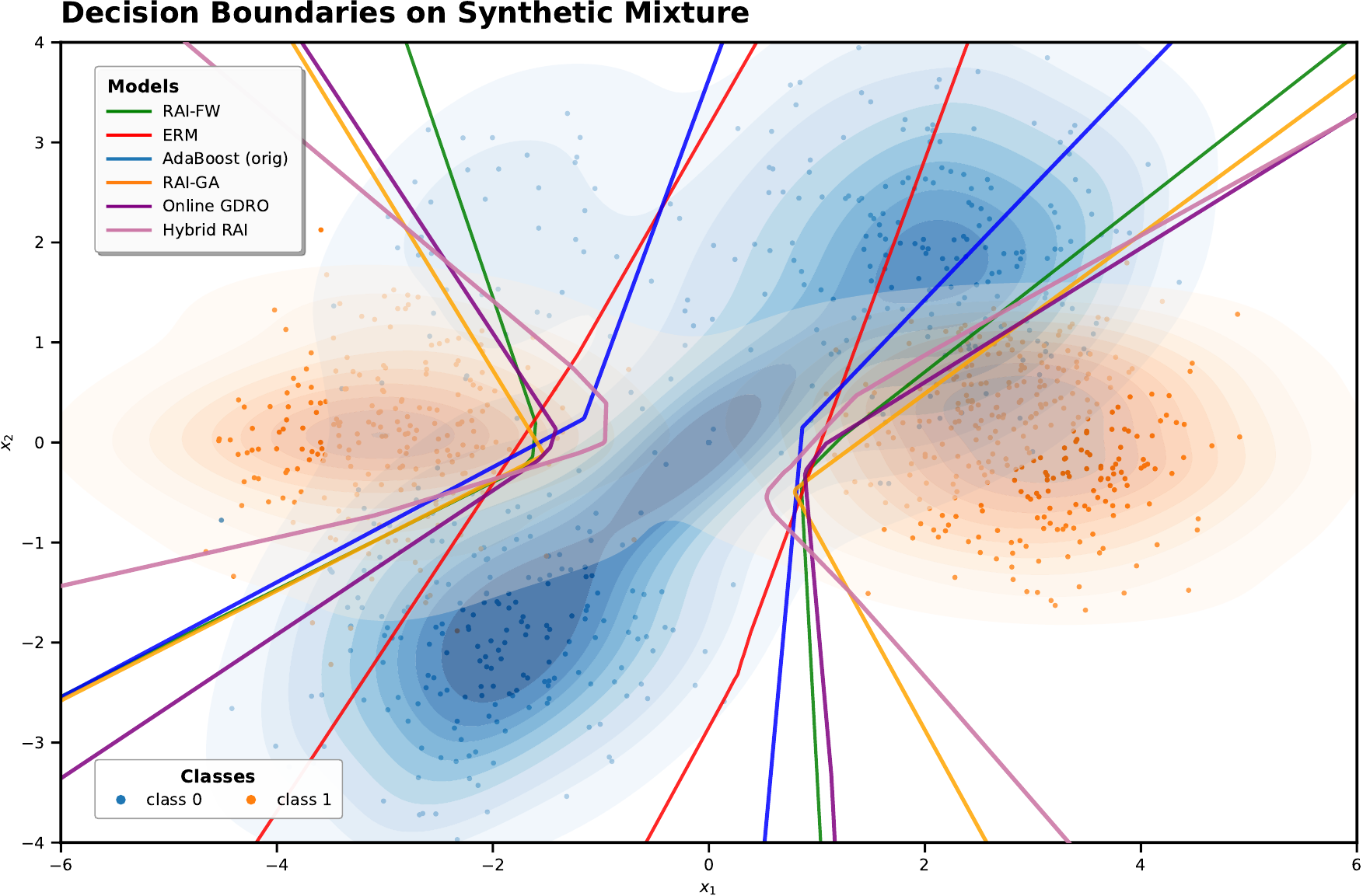}
   % \caption{Imbalanced Logistic Regression Model using Stochastic Gradient Descent.}
    \caption{Decision boundaries of the proposed hybrid RAI model versus baseline algorithms on a synthetic two-class mixture dataset.}
    \label{fig:Baseline}
\end{figure}

After the baseline was established, some performance of the original RAI was specifically noted for its ability to optimize only against one harm model at a time, while the hybrid can use a mixture of models. Also, due to the iterative dynamic adversary $\pi_t$, which is updated each round, it matches non-stationary shifts within the data. For this reason, an iterative testing loop was developed, which included different seeds to change the randomly generated data with different groups, to try and find optimal parameters and compare the performance in both the worst-group accuracy and overall accuracy. 
\begin{comment}

\begin{figure}[h]
    \centering
    \includegraphics[width=.48\textwidth]{xAlgo-Figs/decision_boundaries_comparisons-cropped.pdf}
   % \caption{Imbalanced Logistic Regression Model using Stochastic Gradient Descent.}
    \caption{Closer comparison between fairness models. }
    \label{fig:Comparison}
\end{figure}
\end{comment}
To stress test each agent, we evaluated the standard Empirical Risk Minimization (ERM), RAI-Greedy, and hybrid on a separate audit distribution. A new draw from the same 2-D mixture with increased class overlap was used to emulate adverse operating conditions. Two kinds of service breach were flagged: low-confidence events, which indicate degraded service even if correct, and critical errors, which are wrong predictions. Per agent, the overall accuracy, the worst-group accuracy and the gap, which is the average accuracy subtracted from the worst-group accuracy, are reported. The low-confidence violation rate and the critical error rate are also reported. The end-user report surfaces only the high-level metrics, while the operator report adds per-group accuracies, as well as a table of problem points. This is kept from the user to prevent network information from being leaked from these reports. An agent is marked SLA Violated if the worst group falls below a policy floor, which will be the case due to the complexity of the given dataset. To display these logs in a clean format, a Python front-end package, Streamlit, was used. This allowed multiple figures and tables to be displayed in an interactive manner.

\begin{figure}[h]
    \centering
    \includegraphics[width=.48\textwidth]{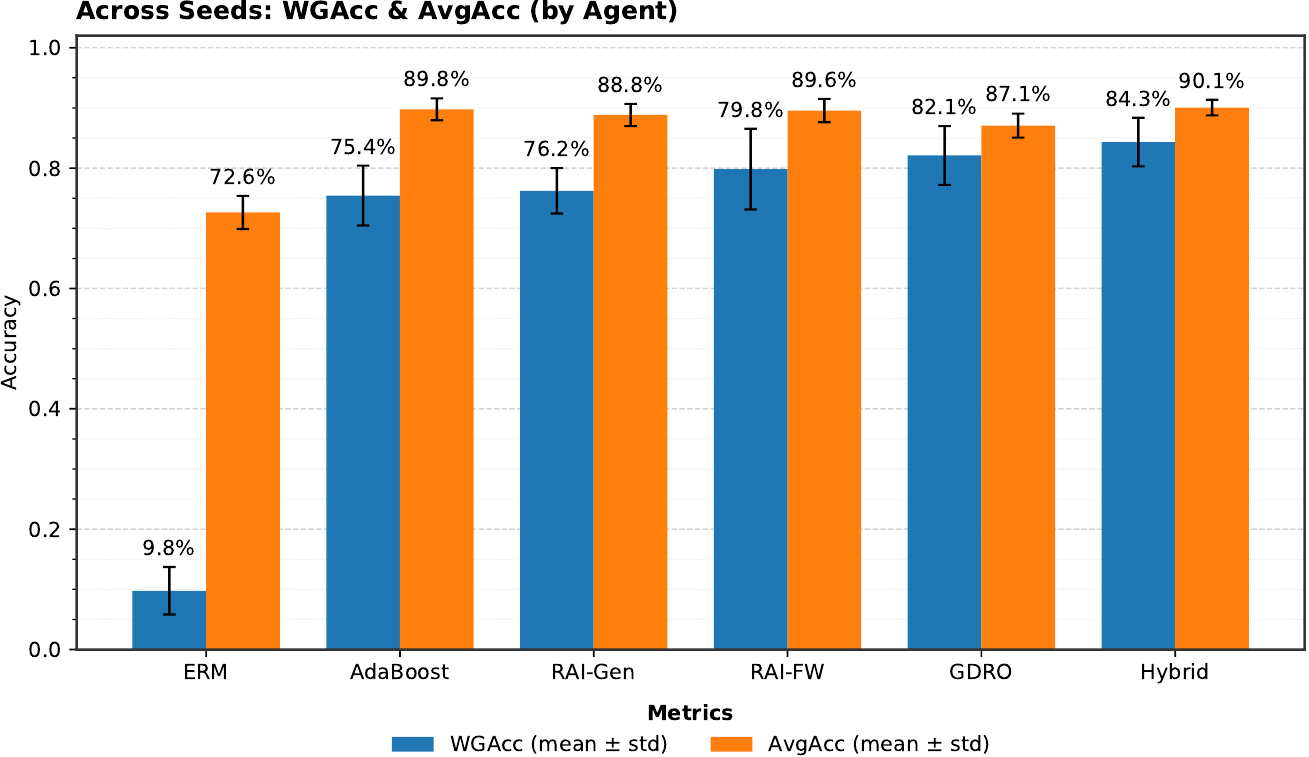}
   % \caption{Imbalanced Logistic Regression Model using Stochastic Gradient Descent.}
    \caption{Worst-group and average accuracy across models and random seeds.}
    \label{fig:wgacc}
\end{figure}

\section{Results and Analysis}

As shown in Figure~\ref{fig:Baseline}, the non-RAI baselines produce near-linear decision boundaries that fail to capture complex overlap between classes. Although they report high overall precision, this is largely due to class~0 dominating the dataset across labels and subgroups. In contrast, both the RAI and hybrid RAI models adaptively reshape the decision boundary, bending it to enclose class~1 regions while pushing it away from class~0. The hybrid variant further refines this separation by trimming class~0 outliers, resulting in a smoother and more balanced margin.

To examine subgroup performance, four representative subgroups were generated from the synthetic dataset. Figure \ref{fig:wgacc} reports mean and standard deviation of worst-group accuracy (WGAcc) and overall accuracy (AvgAcc) across seeds for ERM, AdaBoost, RAI-GA, RAI-FW, GDRO, and the Hybrid. ERM attains a reasonable AvgAcc but collapses on WGAcc, confirming that averages hide subgroup failures. Single-adversary RAI variants raise WGAcc with small to moderate drops in AvgAcc. The Hybrid achieves the highest WGAcc while also matching or exceeding the best AvgAcc among baselines, indicating a superior fairness–utility trade-off with low variance across seeds. Conditional Value at Risk (CVaR) at level $\alpha$ is the average loss over the worst 
$\alpha$ fraction of example. When used as a training objective, CVaR is minimized rather than the mean loss, so updates focus on the hardest cases/subgroups that dominate the tail.

For experimental consistency, all models were trained using identical hyperparameters: 10 total SGD rounds, a batch size of 32, a learning rate of 0.01, a momentum of 0.9, and a weight decay of $5\times10^{-4}$. For the RAI-CVaR model, $\alpha=0.10$, $\eta=0$, and $\lambda=-0.5$ were used. The hybrid RAI used the same CVaR parameter ($\alpha=0.10$) together with stochastic exploration $\epsilon_0=0.05$ and decay factor $\gamma=0.95$. Both methods shared the same inner-game solver to ensure a fair comparison.

\begin{figure}[h]
    \centering
    \includegraphics[width=.48\textwidth]{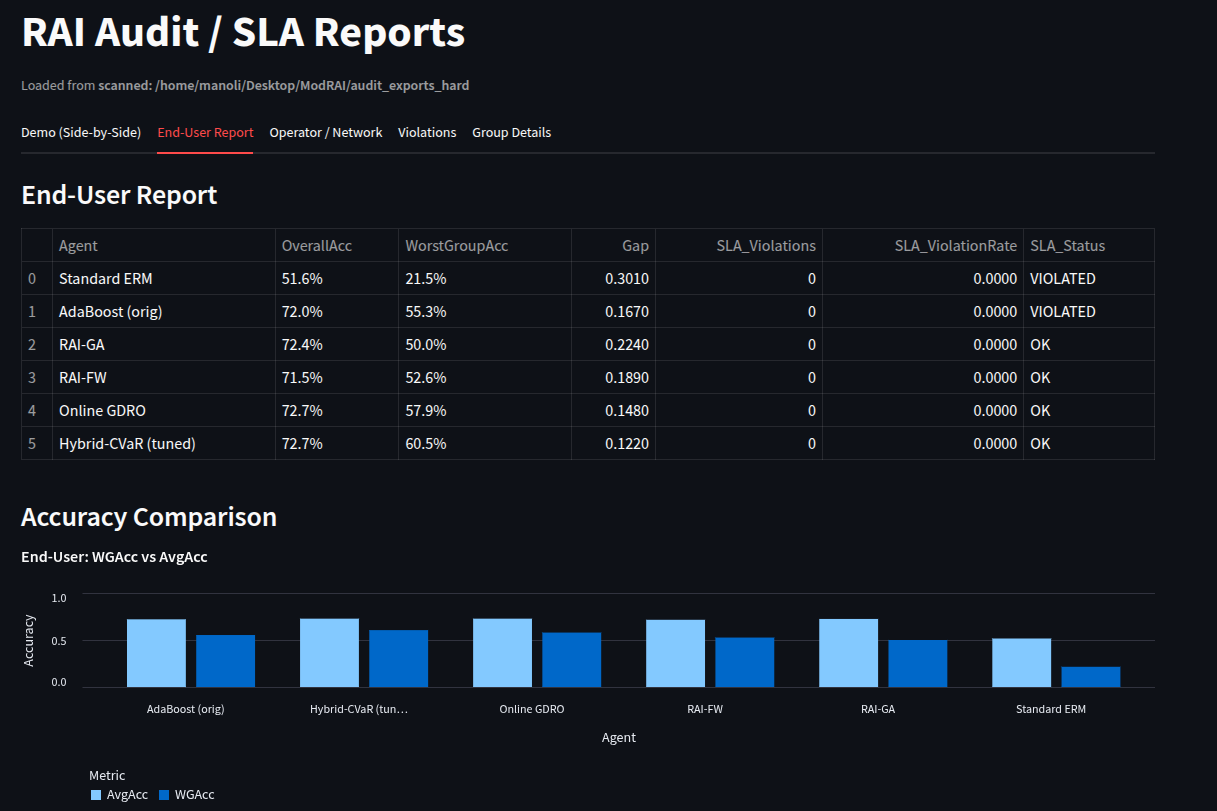}
   % \caption{Imbalanced Logistic Regression Model using Stochastic Gradient Descent.}
    \caption {Comparative end-user showing accuracy, fairness gap, and SLA compliance across Standard ERM, RAI-FW, Adaboost, RAI-GA, Online GDRO, and Hybrid RAI.}
    \label{fig:ereports}
\end{figure}

\begin{figure}[h]
    \centering
    \includegraphics[width=.48\textwidth]{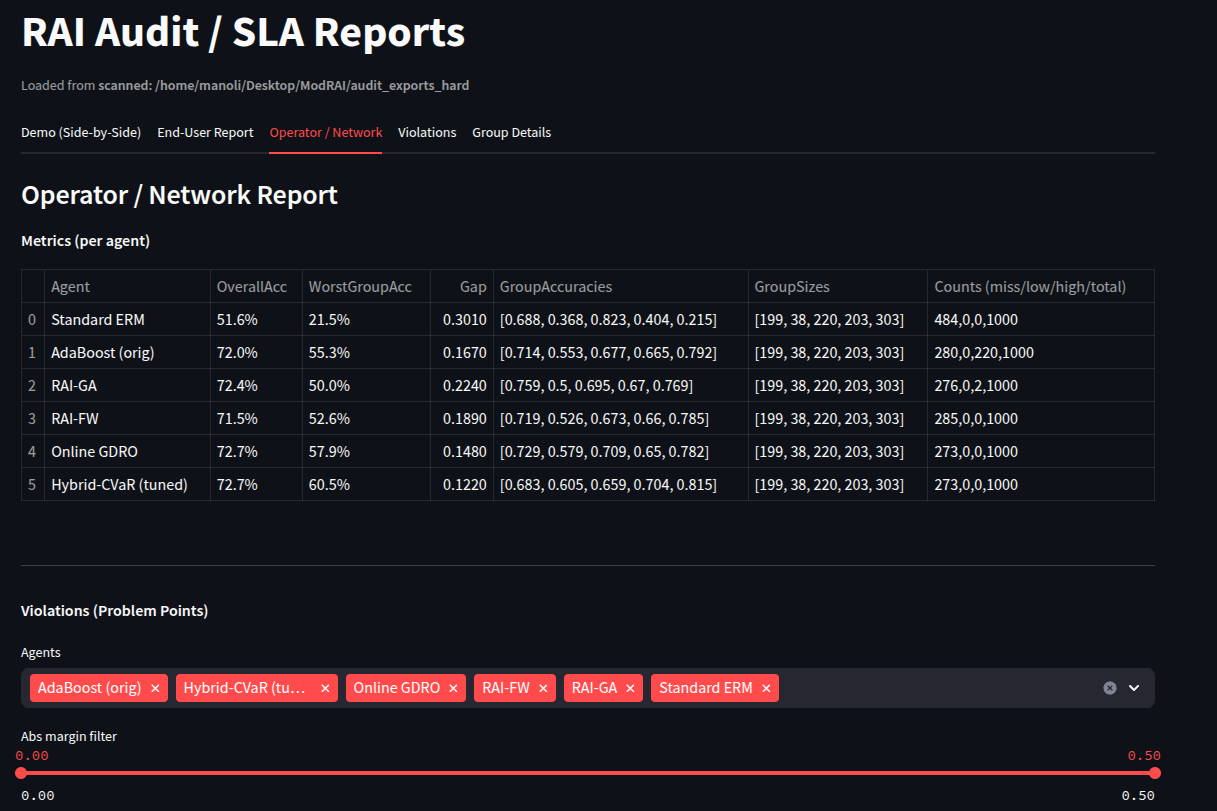}
   % \caption{Imbalanced Logistic Regression Model using Stochastic Gradient Descent.}
    \caption {Comparative end-user showing accuracy, fairness gap, group accuracies, group sizes, and group margins across Standard ERM, RAI-FW, Adaboost, RAI-GA, Online GDRO, and Hybrid RAI.}
    \label{fig:oreports}
\end{figure}

As shown in Figures \ref{fig:wgacc},~\ref{fig:ereports} and~\ref{fig:oreports} in all three figures, the hybrid model consistently outperforms in fairness-critical metrics while maintaining high average accuracy. It achieves the highest AvgAcc and, more importantly, the highest Worst Group Accuracy (WGAcc), thereby reducing the performance disparity relative to both the ERM and RAI-Greedy baselines. Across six agents, Hybrid-CVaR (tuned) attains the best worst-group accuracy (WGAcc) at 60.5\%, a +39.0 pp gain over Standard ERM (21.5\%) and +2.6 pp over the strongest single-adversary baseline, Online GDRO (57.9\%). It also yields the smallest disparity across subgroups (Gap = 0.122 vs. 0.301 for ERM and 0.148 for GDRO), indicating a more uniform performance profile over the five groups (sizes: 199/38/220/203/303). The consistent worst-case subgroup (the small n=38 group) improves from 0.368 (ERM) to 0.605 (Hybrid), while other groups remain competitive (e.g., group-5 at 0.815).

Crucially, the Hybrid matches the top overall accuracy (72.7\%, tied with GDRO), demonstrating fairness gains without sacrificing average utility. Problem-point counts also drop markedly relative to ERM (misses 273 vs. 484), suggesting fewer potential SLA-risk cases. Taken together, the Hybrid offers the most favorable fairness–utility trade-off in this setting—lifting the floor (WGAcc) and narrowing gaps while maintaining state-of-the-art overall accuracy.

The proposed hybrid RAI model demonstrates superior decision boundary behavior due to its integration of Frank–Wolfe optimization with a responsibility-aware regularization framework. The FW algorithm operates through projection-free convex updates, generating smooth and continuous convergence trajectories that naturally align with the underlying data manifold. Gradient-based hybrid RAI updates maintain global consistency in the feature space, which is different from boosting-based models such as AdaBoost that rely on discrete tree partitions and produce fragmented piecewise decision surfaces. This allows the model to adaptively refine its boundary in response to the joint data density, resulting in coherent transitions that effectively capture nonlinear class separations within the synthetic mixture.

RAI adds an uncertainty-aware penalty that discourages overconfident predictions in overlapping regions, promoting calibration and reducing overfitting to local noise. Consequently, the boundary smooths across the central overlap, reflecting balanced treatment of both classes under uncertainty. This convex-plus-stochastic regularization improves generalization and robustness, yielding steadier, smoother decisions than standard ensemble or online-gradient baselines.

Based on the qualitative boundary behavior in Fig.~\ref{fig:Baseline}, the quantitative comparison in Fig.~\ref{fig:wgacc} confirms the robustness and fairness of the proposed hybrid RAI model. The figure shows that Hybrid RAI consistently achieves the highest accuracy in the worst group (WGAcc = 71.4\%) and the average accuracy (AvgAcc = 87.9\%) among all variants tested, including RAI-Greedy CVaR (0.10) and $\chi^2$. These results indicate that the proposed model maintains high overall accuracy and significantly reduces the disparity between performance groups, thus improving fairness and bias mitigation. The improvement arises from the integration of stochastic exploration and dynamic adversarial reweighting within the hybrid optimization process, enabling adaptive learning under non-stationary and imbalanced conditions. Consequently, the hybrid RAI model demonstrates superior stability and equitable performance across all subgroups, outperforming baseline RAI methods that tend to trade off worst-case robustness for marginal gains in mean accuracy.

\section{Conclusion}

This work addressed the responsibility gap that arises when multivendor autonomous 6G networks rely on closed-loop AI for SLA assurance. We proposed a hybrid RAI–Stochastic training objective that preserves the min–max guarantees of RAI while introducing a dynamic adversary and controlled exploration, and we embedded it in a Responsibility-Aware Audit Plane for end-to-end traceability. On a controlled two-feature two-class benchmark with subgroup stresses, the hybrid approach consistently lifted the floor on worst-group performance while retaining strong average accuracy and lowering SLA-violation rates relative to single-adversary RAI baselines. The RAAP also produced actionable accountability indices that localized simulated violations to the responsible agents/vendors, demonstrating that fairness, robustness, and auditability can be operationalized together within CLA workflows. The hybrid approach provides a practical path toward SLA-aligned, accountable automation in multivendor 6G networks and sets the stage for rigorous, practical deployments with measurable guarantees through unifying fairness-aware training with dynamic, exploratory game play and audit-grade logging.

\bibliographystyle{IEEEtran}
\bibliography{xalgo}
\end{document}